# Специфицирование в комплексной САПР реконструкции предприятий на основе модулей в чертеже и электронных каталогов

В. В. Мигунов, канд. техн. наук

***Изложен опыт автоматизации специфицирования проектов реконструкции промышленных предприятий на основе специальных модулей в чертеже, содержащих видимое изображение и дополнительные параметры, и электронных каталогов***

## Введение

Проекты реконструкции промышленных предприятий охватывают много небольших объектов с различных сторон – технологической, монтажной, строительной, электротехнической, санитарно-технической и др. согласно требованиям системы проектной документации для строительства (СПДС). В этих условиях для целей автоматизации приходится набирать несколько различных САПР [1] с естественными сложностями в освоении их интерфейса и неудобством работы в разных средах, отсутствием или недостаточной степенью интероперабельности [2]; задачи специфицирования решаются практически в каждой специализированной САПР, но индивидуальным способом. Обычно проекты выполняются силами небольших проектно-конструкторских подразделений самого предприятия, что делает невозможной узкую специализацию рабочих мест и усиливает эти трудности. В чертежах есть общие части: строительная подоснова, технологическая схема. Они должны передаваться от одного проектировщика к другому и допускать доработку на всех рабочих местах. Еще одна особенность проектов реконструкции – они часто выполняются "поверх" прежних чертежей, хранящихся в бумажном архиве.

Задача автоматизации проектирования реконструкции промышленного предприятия требует программного обеспечения, ориентированного на выпуск чертежей по требованиям ЕСКД и СПДС, работающего одновременно с растровой и векторной графикой, в едином интерфейсе пользователя автоматизирующего подготовку чертежей различных марок СПДС в одном графическом формате.

Теории таких систем практически нет. Из 807 диссертаций по специальности 05.13.12 (САПР), защищавшихся в 1987 – 2003 г.г. с передачей авторефератов в РГБ, ни одна не посвящена этой проблеме (выборка с http://aleph.rsl.ru, 16.08.2004). Данная статья посвящена одной из проблем, решение которых необходимо при создании названных САПР - задаче специфицирования чертежей.

В рамках комплексной САПР реконструкции предприятий естественным образом возникают специализированные проблемно-ориентированные расширения для подготовки чертежей отдельных марок СПДС. В каждом расширении имеются свои структуры хранения данных (параметрические представления), в том числе специфицирующих свойств. В тех частях чертежа, которые не имеют соответствующего специализированного сервиса, также есть элементы, подлежащие специфицированию. При задании специфицирующих свойств необходим единообразный доступ к электронным каталогам соответствующей проблемной ориентации, наряду с возможностью ручного ввода этих свойств. Обычно специфицируется лишь реконструируемая часть, что требует универсальных методов указания областей, подлежащих специфицированию. Сами табличные конструкторские документы (ТКД) отличаются в разных частях проекта составом граф, организацией разделов и др.

Далее описываются модели и методы, позволяющие решить эти вопросы комплексирования САПР реконструкции предприятий. Изложение основано на опыте разработки отечественной САПР TechnoCAD GlassX [3]. Вначале дается понятие "модуля в чертеже", затем описаны построенные на его основе методы подготовки таблиц и задания специфицирующих свойств, и, наконец, специальные решения для комплексирования электронных каталогов.

### Модуль в чертеже

Элемент электронного чертежа типа "Модуль" включает видимую в чертеже совокупность геометрических элементов и невидимое в чертеже параметрическое представление моделируемого объекта. Первично параметрическое представление, связанная с ним геометрическая часть перегенерируется при его изменениях.

Параметрические представления объектов в модулях позволяют реализовать самые разные модели объектов и методы их проектирования [4]. Специализация модулей легко опознается в чертеже. Высокая структурированность параметров обеспечивает быстрый доступ к ним для различной обработки. Легко автоматизируется сбор сведений при генерации спецификаций, при контроле дублирования позиционных обозначений.

Как элемент чертежа модуль подчиняется обычным правилам: его можно удалить, подвинуть, растянуть и т.д., к его геометрическим элементам возможны привязки, он может быть помещен в графическую библиотеку. Все элементы модуля лежат на одном слое, но могут иметь разные типы линий и цвет.

При задании параметрического представления используется как стандартный пользовательский интерфейс (меню, формы ввода...), так и специализированный для черчения и корректировки.

### Табличный модуль

Табличный модуль (ТМО) - модель ТКД, например спецификации, заказной спецификации, монтажной ведомости трубопровода, таблицы колодцев или дождеприемников и т.д.

Параметрическое преставление ТКД включает массив записей, нулевая запись - шапка таблицы. Структура всех записей одинакова и задается последовательными актами деления блоков - прямоугольных фрагментов таблицы на части по вертикали и горизонтали, с признаками видимости этого деления в шапке и в области данных, на фиксированное или произвольное число частей, вплоть до неделимых блоков - клеток таблицы. Таким образом легко создаются ТКД с одинаковой шапкой, имеющие разделы (рубрики) или не имеющие их. После автоматического заполнения ТКД проектировщик создает нужные ему разделы и выполняет автоматизированные операции фасовки строк, упорядочивания по задаваемой последовательности граф, сливания одинаковых и выделения общих частей наименований.

Иерархическая модель ТКД порождает сложности в программировании. Отсутствует понятие строки, и для идентификации выбираемой клетки требуется рекурсивный подсчет высот всех вышестоящих блоков данной записи. Однако эти трудности окупаются "интеллектом" ТМО. Например, при вставке ячеек в таблицу в зависимости от точки указания вставляется именно тот прямоугольный блок, который является частью от деления на произвольное число частей по вертикали. При этом для фланцевых соединений, включающих сразу 5 изделий, вставляются одновременно по одной строке для фланца и прокладки и три строки для шпильки, гайки, шайбы (рис.1).

Для правки ТКД в специальный редактор таблиц, где имеется доступ к электронным каталогам и не допускается иерархия блоков, из табличного модуля передается максимально возможная прямоугольная область, допускающая корректную работу с каталогами. Предусмотрен "буфер изделий", позволяющий переносить группу строк с одноименными специфицирующими свойствами между ТКД, в том числе различных видов (рис.рис.2,3). Эти преимущества достигаются не специализацией кода САПР для конкретных ТКД, а собственно структурой таблиц, задаваемой во внешнем файле, и идентификацией полей ТКД в САПР.

Сервис ТМО дает возможность продолжать ТКД влево или вправо кусками нужной высоты, иметь строки с номерами граф или повторять шапку, менять типы разделительных линий, шрифты текстов в каждой клетке и др. с мгновенной перегенерацией изображения. Комплект параметров табличного модуля можно записать на диск для последующей доработки и использования как прототипа.

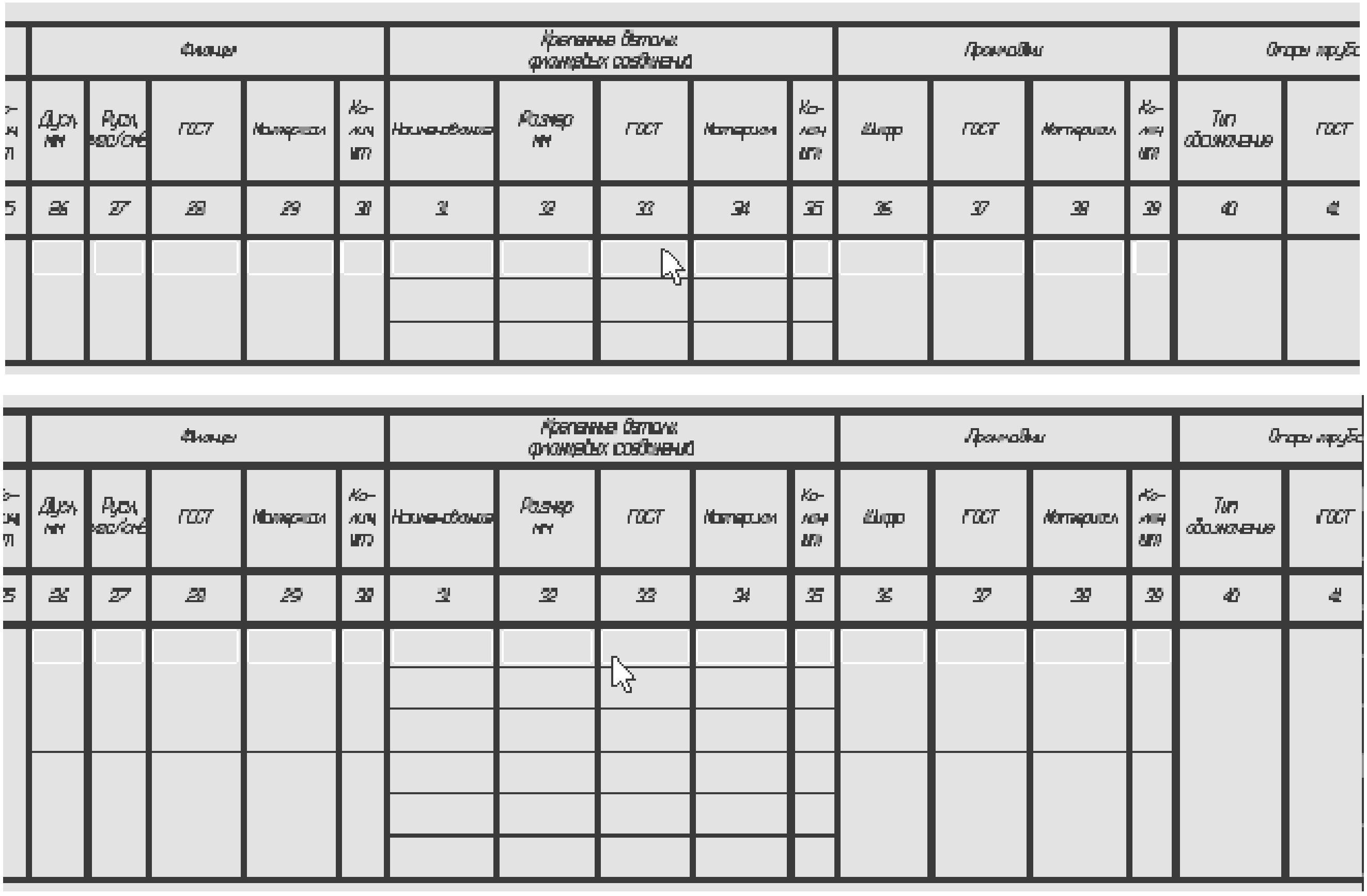

Рис.1. Указание места вставки и вставка блока для специфицирования фланцевого соединения в монтажной ведомости трубопровода

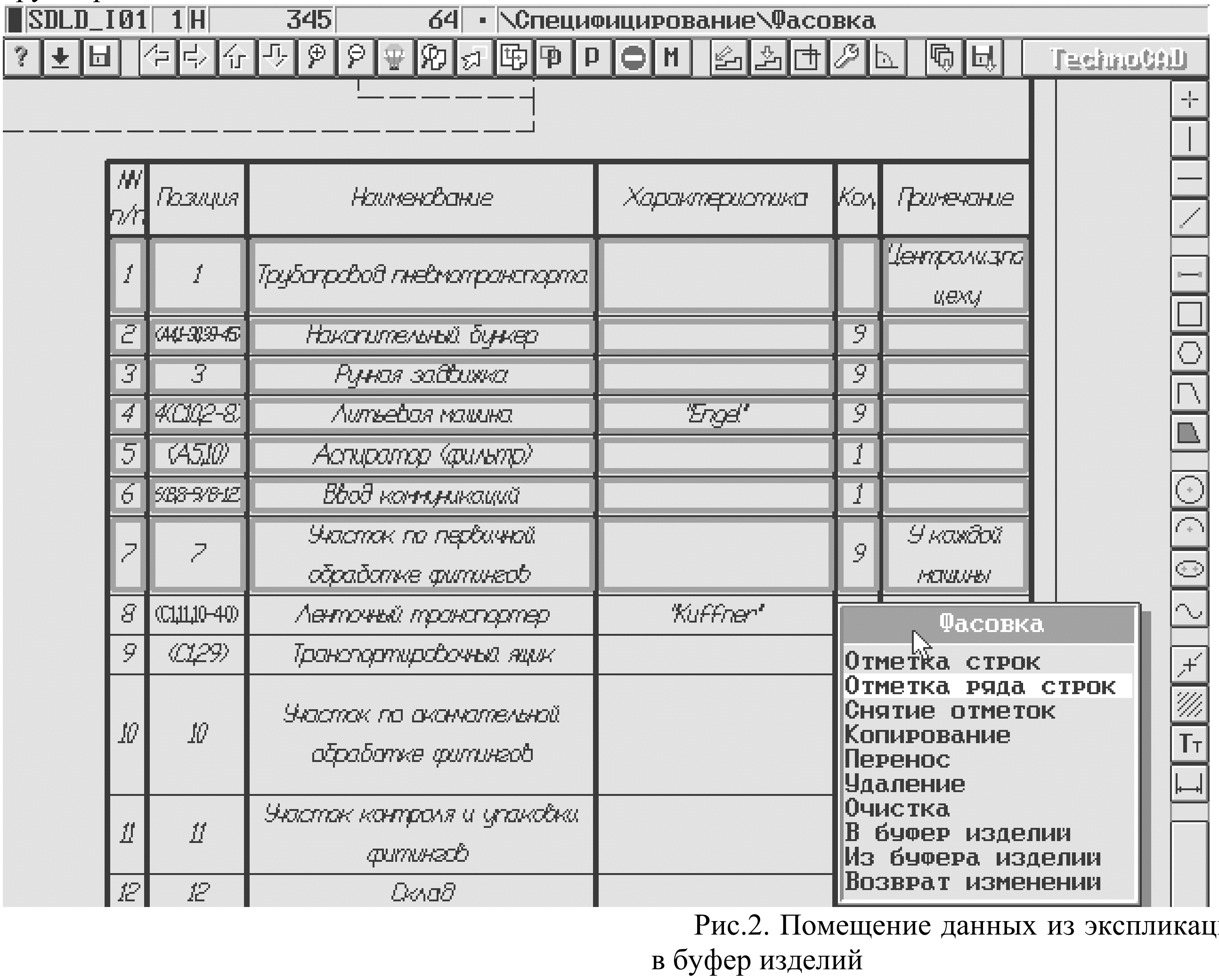

Рис.2. Помещение данных из экспликации в буфер изделий

| Поз. | Обозначение | Наименование | Кол. | Масса ед.кг | Приме-чание |
|---|---|---|---|---|---|
| | | | | | |
| 1 | | Трубопровод пневмотранспорта | | | Центральному цеху |
| [illegible] | | Накопительный бункер | 9 | | |
| 3 | | Ручная задвижка | 9 | | |
| [illegible] | | Литьевая машина | 9 | | |
| (А5,10) | | Аспиратор (фильтр) | 1 | | |
| [illegible] | | Ввод коммуникаций | 1 | | |
| 7 | | Участок по первичной обработке фитингов | 9 | | У каждой машины |

Рис.3. Результат вставки из буфера изделий в спецификацию

**Позиционные обозначения**

Модуль типа "Позиционное обозначение" (ПО) используется для задания специфицирующих свойств произвольным фрагментам любого чертежа (изделиям), причем с самими этими фрагментами он не связывается. Модуль ПО всегда содержит один видимый многострочный текст, понимаемый как обозначение(я) одного или нескольких изделий в зависимости от значения свойства "Тип ПО", и невидимые на чертеже специфицирующие свойства. Он может быть создан заново или получен преобразованием имеющегося в чертеже текста.

Состав специфицирующих свойств является объединением всех полей из спецификаций по ГОСТ 21.101-97 и заказных спецификаций по ГОСТ 21.110-95: "Марка, Поз.", "Позиция", "Обозначение", "Наименование", "Количество"; "Масса ед."; "Примечание", "Тип, марка", "Наименование и техническая характеристика", "ЕдИзм", "Код оборудования", "Завод-изготовитель".

Специфицирующие свойства одного или нескольких изделий модуля ПО хранятся как таблица свойств и правятся в специальном редакторе таблиц. Возможна правка специфицирующих свойств сразу для нескольких выбираемых модулей ПО. В редакторе таблиц осуществляется выбор в электронных номенклатурных каталогах (ЭНК), возможно групповое специфицирование узлов (например, комплекта деталей фланцевого соединения).

При добавлении ПО производится автоматический контроль дублирования в соответствии с текущей установкой области контроля, которая может задаваться из трех частей: модули ПО, модули "Аксонометрическая схема", модули "Профиль наружной сети ВК". При добавлении дублирующего ПО дается сообщение о повторении и запрос подтверждения помещения его в чертеж. Имеется операция контроля дублирования ПО внутри группы чертежей на диске.

Для лучшей ориентации в уже использованных обозначениях и их местонахождении в чертеже можно просмотреть их список и провести поиск строки в текущем чертеже.

У модулей ПО есть дополнительное свойство "Тип объекта позиционного обозначения". Оно относится к изделию, на которое указывает ПО, и имеет три варианта: "Нет типа", "Труба", "Колодец" и используется следующим образом:

если задан тип объекта "Труба", то при выборе в ЭНК их список автоматически сужается, в нем остаются только трубы;

если задан тип объекта "Колодец", то позиционное обозначение учитывается при автозаполнении таблиц колодцев и не учитывается при автозаполнении спецификаций и заказных спецификаций, для других типов объектов - наоборот.

В результате анализа нескольких сот чертежей в составе документации на отечественные и зарубежные технологические процессы и линии выявлено наличие самых различных систем алфавитно-цифровых ПО. Единственным общим моментом для различных систем ПО является серийно-порядковый способ их структурирования, с разделителями серий или без них.

Для целей автоматического упорядочивания строк спецификаций, анализа корректности и дублирования ПО в чертежах разработаны правила упорядочивания и структурирования ПО. Позиционные обозначения разбиваются на структурные части слева направо с разделителями из набора ".-/()", упорядочиваются по возрастанию 1й части, внутри группы ПО с одинаковой 1й частью упорядочиваются по возрастанию 2й части и т.д. Упорядочивание по возрастанию означает, что числа идут раньше букв, арабские числа раньше римских, все числа по возрастанию значений, русские буквы раньше латинских, прописные буквы раньше строчных, в алфавитном порядке русского и английского языков.

На основе правил упорядочивания автоматически строятся списки ПО или их структур с частотами. Список структур ПО одного чертежа позволяет выявить неверную замену символов русский/английский, ноль/буква "O", отсутствие разделителя или неверный разделитель.

**Электронные каталоги**

Электронные номенклатурные каталоги выпускаемых изделий получают все большее распространение в области маркетинга, продаж и сопровождения продукции. Здесь ЭНК ускоряют выбор и заказ товара в магазине, в автосервисе, в Internet. В сфере проектирования роль каталогов существенно иная: анализ технических характеристик, принятие решений и описание их в проектно-сметной документации. В частности, в задачах реконструкции предприятий такими описаниями можно считать заказные спецификации, где собираются сведения для осуществления поставки изделий, и чертежи со спецификациями для проведения строительно-монтажных работ. В комплексной САПР реконструкции предприятия необходимы ЭНК, охватывающие различные части проекта. В рамках одной части проекта встречаются варианты объединения ЭНК (например, [5, 6]), но они в основном представляют собой электронные версии бумажных каталогов и не ориентированы на автоматизированное специфицирование по СПДС.

В TechnoCAD GlassX разработаны системотехнические решения, позволившие в рамках одной САПР эффективно решить задачу работы с ЭНК при подготовке чертежей марок МТ (монтажно-технологическая часть, профиль работ МТ); А.. (автоматизация, профиль работ КИП); ОВ, ВК, НВК, ТС (отопление, вентиляция, водоснабжение, канализация, теплоснабжение, профиль работ ОВК).

Каталоги включают сведения из ГОСТов, ОСТов, ТУ, государственных реестров, каталогов производителей оборудования, труб, арматуры, элементов трубопроводов, приборов и исполнительных механизмов. ЭНК ориентирован на сложившуюся практику крупного предприятия и не претендует на полноту, но является достаточно представительным для выявления типовых и особенных свойств каталогов для целей автоматизированного проектирования (строка 1 таблицы).

Таблица. Структурно-количественные характеристики ЭНК

| | | | | |
|---|---|---|---|---|
| 1 | Профиль работ (часть проекта) | МТ | КИП | ОВК |
| 2 | Бумажных каталогов /таблиц | 111 /110 | 14 /52 | 102 /103 |
| 3 | Наименований свойств изделий | 154 | 250 | 232 |
| 4 | Свойств, измеряемых в мм /без наименования | 87 /38 | 9 /4 | 143 /57 |
| 5 | Строк в таблицах всего/min /max | 39912 /1 /9726 | 4228 /2 /933 | 24073 /1 /9726 |

Не каждый бумажный каталог содержит графическую информацию. Если она имеется, то служит в основном для задания смысла и обозначения размеров, приводимых в табличной или текстовой частях каталога. Количество свойств, измеряемых в миллиметрах, составляет около 60% для профилей МТ и ОВК, тогда как для КИП - менее 4% (строки 2, 3 таблицы). В большинстве случаев эти величины имеют достаточно прозрачные словарные наименования, и обращаться к их обозначениям в чертежах нет необходимости.

Даже при отсутствии ясных наименований для проектировщика, много лет работающего в своей сфере, обозначения размеров становятся привычными и не требуют работы с графической частью каталогов. Принятое решение отказаться от графических частей в ЭНК не вызвало нареканий со стороны проектировщиков.

Текстовые и табличные части бумажных каталогов всех трех профилей работ однотипны и несут основную информацию, необходимую проектировщику. В текстовую часть (включая наименование самого каталога) выносятся общие, единые для каталога сведения об областях применимости изделий, такие как назначение, агрессивность сред, диапазоны температур и давлений, варианты материалов. Здесь же помещаются правила оформления обозначения и заказа, то есть структура наименования изделия в заказной спецификации обозначения в спецификации. Часть сведений в тексте может относиться не ко всем, а к некоторым изделиям.

Наиболее структурированная, табличная часть информации содержит допустимые варианты сочетания параметров изделий. При этом в одной ячейке таблицы могут встретиться несколько вариантов значений параметров.

Для всех профилей работ удалось создать единую информационную структуру на основе таблиц в DBF формате и текстовых файлов с правилами. В основном каждая таблица соответствует одному бумажному каталогу, но иногда удобнее слить два похожих каталога в одну таблицу, или разбить каталог на две таблицы. Столбцы таблиц именуются универсально: MARKA, X_1, X_2, X_3… и хранят исходные данные, из которых при занесении изделия в конструкторский документ автоматически генерируются марки, обозначения, наименования и другие свойства.

Имеется метатаблица с описанием структур таблиц, включающим русское наименование или обозначение, единицы измерения, тип данных для всех столбцов каждой структуры. Единицы измерения в ЭНК сохраняются те же, что были в бумажных каталогах. Еще одна спецтаблица для каждой из таблиц с данными задает имя ее структуры, наименование таблицы и породившего ее каталога, ее классификационные признаки (служат для сокращения вариантов при выборе). В ячейках таблицы с данными находятся характеристики изделий не только числовые, но и, например, "для воды, пара, масла, нефти и жидких нефтепродуктов", "Георгиевский арматурный завод", "ТУ 26-07-1440-87". Также здесь встречаются "непосредственные меню" - списки допустимых вариантов, выбор из которых предлагается проектировщику тогда, когда он уже остановился на каком-либо изделии.

Когда проектировщик выбрал изделие, все значения в строке таблицы получают конкретные значения (не списки вариантов). Далее из этих значений генерируются сведения для заполнения полей в спецификациях по правилам, указанным в файле правил и полностью соответствующим заданным в исходном бумажном каталоге. Правила для каждой таблицы с данными и для каждого поля спецификаций задают сложение фрагментов строк. Каждый фрагмент может быть константой, значением поля в таблице данных, результатом выбора из меню (рис.4, списки опций "внешних" меню хранятся во внешнем файле и идентифицируются именем меню), результатом ввода проектировщиком числового или строкового значения, а также временной переменной, в которой сохранено введенное, выбранное или сгенерированное ранее (для другого поля спецификации) значение. Кроме полей заказной и обычной спецификаций, таким же образом генерируются сведения и для некоторых специальных задач в САПР: наружный диаметр трубы для указания на профиле наружной сети водоснабжения и канализации, материал и др. Все эти сведения передаются в ядро САПР, где при необходимости числовые параметры приводятся к нужным единицам измерения.

ЭНК реализованы почти полностью в рамках изложенной системной организации. Исключение - иерархически связанные меню выбора защитных гильз для датчиков ТСП/ТСМ и ТХА/ТХК, выбора диапазонов измерений для ДИСК-250 (профиль КИП). Из-за малого количества таких меню их проще было реализовать непосредственно в программном коде САПР, но вызов их осуществляется по командам из файла правил ЭНК.

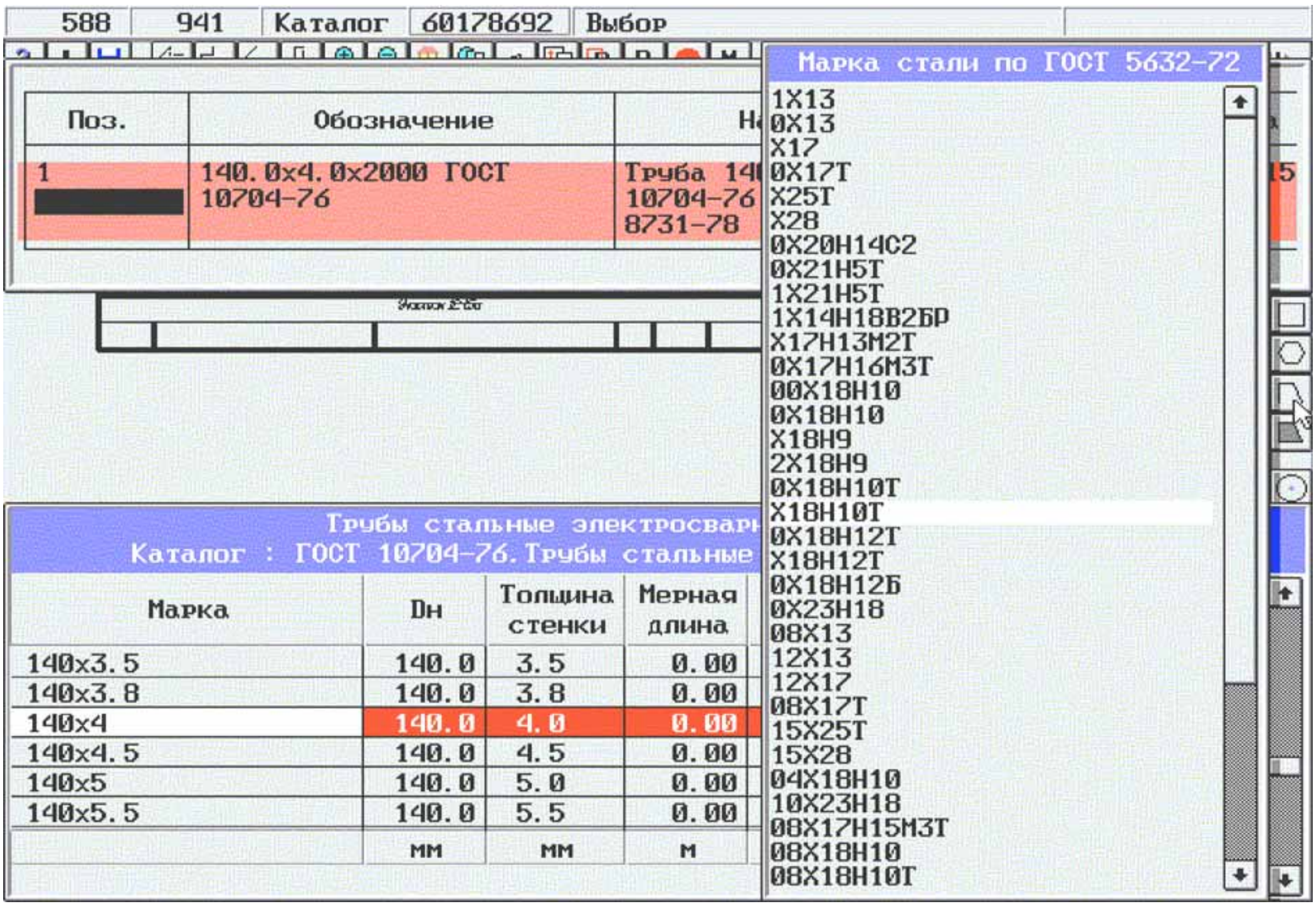

Рис.4. Выбор труб в электронных каталогах

Применение "непосредственных" и "внешних" меню привело к тому, что таблицы с данными для некоторых каталогов содержат одну или две строки; хотя в некоторых случаях не удалось избежать тиражирования строк, но общее число строк во всех таблицах имеет разумные значения (строка 4 таблицы). Вопрос быстрого роста числа строк связан с количеством необходимых столбцов в таблицах ("проклятие размерности"). Как показал анализ, для трех профилей работ лишь малая часть таблиц имеет число столбцов больше 16, и при этих количествах столбцов число строк еще позволяет работать. Дело в том, что реально в ЭНК присутствуют не только сведения, помещаемые в специфицирующие документы, но и многие другие, необходимые для корректного выбора. У приборов для измерения давления в ЭНК имеются характеристики: марка, наименование, назначение (агрессивность среды, жидкая или газообразная, измерение избыточного давления, отсчет разности давлений, ...), класс точности, ГОСТ, ТУ, функциональный признак (PIS, PI), верхние значения диапазона измерений, диаметр корпуса, наличие фланца (задний, передний, нет), расположение штуцера (радиальный, осевой), исполнение и ресурс срабатываний сигнализирующего устройства, примечание (ограничения напряжений, токов, мощностей).

Таблицы с данными имеют четыре варианта классификационных признаков, используемых для ускорения выбора. Многие таблицы КИП разделяются по тому, являются ли изделия первичными или вторичными приборами, и по символам, означающим измеряемую величину (F - расход, P - давление, T - температура и др.). Для профилей МТ и ОВК используется иерархический список принадлежности таблиц группам изделий: "Оборудование/Насос", "Деталь/Тройник" (ключевые слова). Также для МТ и ОВК удобны классификации в виде ограничений по интервалам температур, давлений, диаметров условного прохода, наружных диаметров, диаметров резьбы.

***Литература***